\documentclass[a4paper,fleqn]{cas-dc}

\usepackage[numbers]{natbib}

\usepackage{graphicx}
    \graphicspath{{figures/}}

\usepackage{siunitx}

\def\tsc#1{\csdef{#1}{\textsc{\lowercase{#1}}\xspace}}
\tsc{WGM}
\tsc{QE}
\tsc{EP}
\tsc{PMS}
\tsc{BEC}
\tsc{DE}

\begin{document}

\let\WriteBookmarks\relax
\def\floatpagepagefraction{1}
\def\textpagefraction{.001}

\shorttitle{Fast Stop-Band Evaluation}
\shortauthors{A. Hvatov et~al.}

\title[mode = title]{General Method for Evaluation of Stop-Bands of Periodic Structures with Symmetric Unit Cells}        

\author[1]{Alexander Hvatov}[orcid=0000-0002-5222-583X]

\author[2,3]{Mariia Krasikova}[orcid=0000-0002-5950-4807]
    \ead{mariia.krasikova@metalab.ifmo.ru}
    \cortext[cor1]{Corresponding author}

\author[2]{Aleksandra Pavliuk}[orcid=0000-0001-5277-4313]


\author[3]{Steffen Marburg}[orcid=0000-0001-6464-8015]

\affiliation[1]{
    organization={NSS Lab, ITMO University},
    postcode={197101}, 
    city={St. Petersburg},
    country={Russia}
}

\affiliation[2]{
    organization={School of Physics and Engineering, ITMO University},
    city={St. Petersburg},
    postcode={197101}, 
    country={Russia}
}

\affiliation[3]{
    organization={Chair of Vibroacoustics of Vehicles and Machines, Technical University of Munich},
    city={Garching b. M\"unchen},
    postcode={85748}, 
    country={Germany}
}

\begin{abstract}
The mirror symmetries of a periodic unit cell are exploited to decompose the standing-wave eigenproblem at the high-symmetry vertices of the Brillouin zone into four independent sub-problems on a quarter-cell, each governed by Neumann (sound-hard) or Dirichlet (sound-soft) boundary conditions. Sorting and pairing the resulting eigenfrequencies by index along each segment of the irreducible Brillouin zone boundary yields an explicit formula for the stop-band intervals without computing the full dispersion diagram. The decomposition is exact, following directly from the representation theory of the little group at each high-symmetry point. It applies to any unit cell whose material distribution is invariant under the mirrors normal to the cell faces. The method is validated on two configurations: a phononic crystal of lead cylinders in an epoxy matrix, analyzed using the plane-wave expansion, and a lattice of coupled C-shaped Helmholtz resonators, analyzed using finite-element analysis. For both systems, the reconstructed stop-band boundaries agree with the full Floquet dispersion calculation to within 1\% for the lowest bands, requiring eigenvalue solutions at only three discrete wavevectors. Avoided crossings within a Brillouin zone segment can cause bands to exhibit non-monotone behavior, rendering the pairing rule approximate; the spectral conditions for this are identified. Flat bands common to both boundary-condition types are identified as bound states in the continuum.
\end{abstract}




\maketitle

\section{Introduction}

Periodic structures allow control of wave propagation by tuning pass- and stop-bands, forming the basis for modern wave-engineering applications. A wide range of practical applications benefiting from complex band structures includes noise and vibration insulators~\cite{gao2022acoustic,dalela2022review}, waveguides~\cite{meng2021optical,oudich2023tailoring}, and sensors~\cite{lucklum2009phononic,nair2010photonic,fenzl2014photonic}. Several works also demonstrate multi-purpose structures capable of simultaneous acoustic and elastic wave mitigation~\cite{chen2021hybrid,zeng2024nonlocal,xiao2025simultaneous}, or that are simultaneously phononic and photonic crystals~\cite{sadat-saleh2009tailoring,pennec2010simultaneous,mohammadi2010simultaneous}. 

The analysis of periodic structures rests on the Floquet-Bloch framework~\cite{floquet1883sur,bloch1928uber} that led to the theory of photonic and phononic crystals~\cite{joannopoulos2008photonic, khelif2016phononic}. The Floquet-Bloch framework also reduces wave propagation problems in infinite periodic domains to eigenvalue problems on a single unit cell or a Brillouin zone~\cite{brillouin1953wave}. Rigorous band structure computations can be performed using the plane-wave expansion (PWE) method~\cite{ho1990existence,kushwaha1994theory}, which, however, might suffer from slow convergence related to an inappropriate formulation of the eigenproblem~\cite{cao2004convergence}. The alternative approach is the finite element method (FEM), which offers the flexibility to handle arbitrary unit cell geometries, complex material distributions, and anisotropic media.

In all these approaches, the dispersion relations $\omega_n(\mathbf{k})$ are computed by sweeping the wavevector~$\mathbf{k}$ along the boundary of the irreducible Brillouin zone. The sweep is computationally inexpensive for simple geometries. At the same time, for complex unit cells requiring small discretization steps, the cost per wavevector can be substantial, and tens or even hundreds of $\mathbf{k}$-points are typically needed to resolve the dispersion diagram. Several reduced-order strategies have been proposed to mitigate this cost. For example, reduced Bloch mode expansion (RBME) reduces computation time by selecting a specific set of Bloch eigenvectors at high symmetry points~\cite{hussein2009reduced}. Dispersion relations can also be extracted from mass and stiffness matrices with periodicity conditions using the wave finite element method (WFEM)~\cite{mace2005finite,mace2008modelling}. Efficient computational schemes can also be based on Dirichlet-to-Neumann maps~\cite{yuan2006photonic}. However, all of these approaches require computation over the entire Brillouin zone. A complementary approach bypasses the dispersion calculation and determines stop-band boundaries directly from the eigenfrequencies of a single unit cell.

The rigorous foundation of this property was established after Mead \cite{mead1975wave,mead1975waveII} through the bi-orthogonality relation for free waves in a waveguide \cite{ledet2019bi,hvatov2015free}: the boundaries between pass- and stop-bands are given exactly by the eigenfrequencies of a symmetric unit periodicity cell with 'class-consistent' (Class A and Class B) boundary conditions, a formulation converted to the finite element format and applied to waveguides of arbitrary cross-sectional complexity in \cite{sorokin2022finite}. For axial waves and for the scalar acoustic problems considered in this work, the Class A and Class B conditions collapse to the Dirichlet (sound-soft) and Neumann (sound-hard) conditions. Further generalization allows the treatment of radially-periodic~\cite{hvatov2018application} and nonlinear structures~\cite{hvatov2019symmetrical}, as well as the development of reduced-order models for periodic elastic layers~\cite{hvatov2019assessment}.

Following works on the classification of Bloch modes based on the system symmetry~\cite{bouckaert1936theory,sakoda1995symmetry,sakoda2005optical,joannopoulos2008photonic}, at band edges, Bloch waves become standing waves with a specific parity: one mode with field maxima at the symmetry planes (corresponding to Neumann boundary conditions) and one with nodes (corresponding to Dirichlet conditions). In the general elastodynamic setting, these two families of band-edge standing waves correspond to the Class A and Class B (class-consistent) boundary conditions derived from the bi-orthogonality relation \cite{sorokin2022finite, ledet2019bi}; in the scalar acoustic case treated here, this classification reduces to the Dirichlet/Neumann pair. The frequency separation between these two standing-wave solutions equals the width of the band gap.

Closely related to the symmetry classification of band-edge modes is the recently discovered topological origin of truncation resonances, i.e., eigenfrequencies that appear within band gaps of finite periodic structures. These in-gap resonances are controlled by the Zak phase (a one-dimensional topological invariant) and a phason parameter describing the choice of unit cell termination~\cite{rosa2023material}.

The phenomenon of bound states in the continuum (BIC)~\cite{hsu2016bound,he2025theoretical}, that is, localized modes existing within the radiation continuum, represents another manifestation of symmetry-protected wave trapping in periodic systems. Symmetry-protected BICs arise when a bound mode belongs to a symmetry class orthogonal to all radiation channels, a mechanism directly related to the Dirichlet/Neumann decomposition exploited in the present work.

However, this body of work has been limited to one-dimensional periodic structures. The approach based on modal analysis of a symmetric unit periodicity cell with class-consistent boundary conditions \cite{hvatov2015free,sorokin2022finite} was originally formulated for periodic structures in one spatial direction, where the Brillouin zone boundary consists of only two points and the transfer matrix is finite-dimensional. For two- and three-dimensional periodic media, such as phononic and photonic crystals or periodic metastructures, the Brillouin zone boundary is a continuous path connecting multiple high-symmetry vertices, and the unit cell is a two- or three-dimensional domain. The relationship between boundary conditions and band-edge symmetry is governed by the full point group of the lattice rather than a simple reflection.

Following the approach developed in Ref.~\cite{hvatov2015free}, the present work shows that the mirror symmetries $\sigma_x$ and $\sigma_y$ of the unit cell enable the decomposition of standing-wave eigenstates at Brillouin zone vertices into four independent sub-problems on a quarter-cell, each governed by class-consistent boundary conditions, which in the acoustic case treated here take the Neumann (sound-hard) or Dirichlet (sound-soft) form on each face. The resulting eigenfrequencies, sorted and paired by index across the Brillouin zone boundary segments, reconstruct the stop-band map, as described below. The decomposition is exact, as it follows directly from the representation theory of the little group at each high-symmetry point, and requires only that the material distribution be invariant under the mirrors normal to the cell faces. To demonstrate the method's generality, two configurations of different physical character are examined: a lead/epoxy phononic crystal analyzed using the plane-wave expansion and a lattice of coupled C-shaped Helmholtz resonators analyzed using finite elements. The method is additionally applied to a photonic crystal.

The remainder of this paper is organized as follows. Section~\ref{sec:theory} develops standing-wave decomposition at Brillouin zone boundaries and derives the spectral pairing rule that connects unit cell eigenfrequencies to stop-band boundaries. Section~\ref{sec:methods} describes the computational methods, including the PWE for the phononic crystal and the FEM for the Helmholtz resonator array. Section~\ref{sec:results} presents the results for both configurations and discusses the accuracy and limitations of the reconstruction procedure. Section~\ref{sec:conclusion} concludes with a summary and outlook.

\section{Methods and Materials}

\subsection{Problem formulation}
\label{sec:formulation}

Since the present formulation is restricted to scalar pressure fields, the class-consistent boundary conditions of the general theory \cite{sorokin2022finite} take throughout this paper their acoustic form: Dirichlet (sound-soft) and Neumann (sound-hard).

We recall the standard Floquet–Bloch formulation and consider a two-dimensional acoustic medium periodic with respect to a square Bravais lattice with primitive vectors $\mathbf{a}_1 = a\hat{\mathbf{x}}$, $\mathbf{a}_2 = a\hat{\mathbf{y}}$, and unit cell $\Omega = [-a/2,\, a/2]^2$. The pressure field satisfies the equation

\begin{equation}
  \nabla \cdot \bigl(\sigma(\mathbf{r})\,\nabla p\bigr)
  + \omega^2\,\tau(\mathbf{r})\,p = 0,
\label{eq:wave}
\end{equation}

where $\sigma(\mathbf{r}) = 1/\rho(\mathbf{r})$ is the inverse mass density and $\tau(\mathbf{r}) = 1/\bigl(\rho(\mathbf{r})\,c(\mathbf{r})^2\bigr)$ is the inverse bulk modulus, both periodic with the lattice. By Bloch's theorem \cite{floquet1883sur,bloch1928uber,brillouin1953wave}, the eigenstates can be sought in the form $p_{\mathbf{k}}(\mathbf{r}) = e^{i\mathbf{k}\cdot\mathbf{r}}\,u_{\mathbf{k}}(\mathbf{r})$ with $u_{\mathbf{k}}$ lattice-periodic, which reduces the problem on the infinite domain to an eigenvalue problem on a single cell. Equivalently, one may work directly with the pressure field~$p$ and impose the Bloch boundary condition on the cell faces. Denoting the pair of opposite faces normal to~$\mathbf{a}_j$ as $\partial\Omega_j^{\pm}$ , and introducing the Bloch phase factor $\lambda_j(\mathbf{k}) = e^{i\mathbf{k}\cdot\mathbf{a}_j}$, the condition reads

\begin{equation}
  p\big|_{\partial\Omega_j^+}
  = \lambda_j\,p\big|_{\partial\Omega_j^-},
  \qquad
  \sigma\,\frac{\partial p}{\partial n}\bigg|_{\partial\Omega_j^+}
  = \lambda_j\,\sigma\,\frac{\partial p}{\partial n}\bigg|_{\partial\Omega_j^-}.
\label{eq:bloch_bc}
\end{equation}

For each wavevector $\mathbf{k}$ in the first Brillouin zone, Eq.~\eqref{eq:wave} has a discrete spectrum $0 \leq \omega_1(\mathbf{k}) \leq \omega_2(\mathbf{k}) \leq \cdots\,$, and the functions $\omega_n(\mathbf{k})$ constitute the band structure. The phase factors~$\lambda_j$ are complex-valued, such that the real and imaginary parts of~$p$ are coupled via the boundary conditions.

\subsection{Standing-wave decomposition and the spectral pairing rule}
\label{sec:theory}

Two ingredients of the construction below are established results. First, for waveguides periodic in one direction, the coincidence of pass-/stop-band boundaries with the eigenfrequencies of a symmetric unit periodicity cell under class-consistent boundary conditions \cite{mead1975wave,mead1975waveII,hvatov2015free,ledet2019bi,sorokin2022finite}. Second, the parity classification of Bloch standing waves at high-symmetry points of the Brillouin zone \cite{bouckaert1936theory,sakoda1995symmetry,sakoda2005optical}. The new elements of the present work are: (i) the extension to media periodic in two spatial directions, in which the decomposition is performed independently along each lattice direction and yields four parity sectors on a quarter-cell; (ii) the index-pairing rule, Eq.~\eqref{eq:pairing}, reconstructing partial and omnidirectional stop-bands segment-by-segment along the irreducible Brillouin zone boundary; and (iii) the spectral-isolation criterion $\delta_d/g_n$ identifying when the rule is exact.

\begin{figure*}[htbp!]
    \centering
    \includegraphics[width=0.9\linewidth]{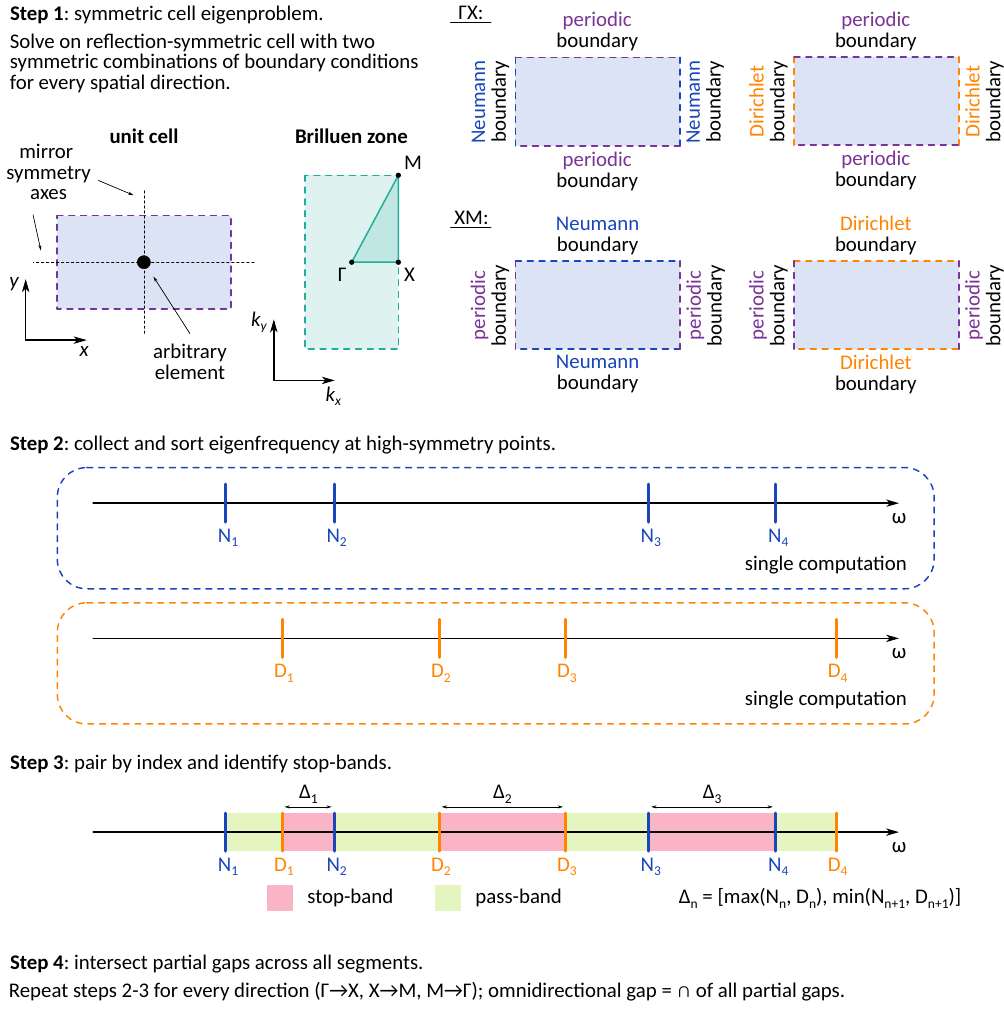}
    \caption{The developed fast stop-band evaluation approach. It reduces the number of computations from the square of the number of frequency discretization points to two.}
    \label{fig:approach}
\end{figure*}

Building on the standard Floquet-Bloch framework recalled above, mirror symmetries of the unit cell reduce the standing-wave eigenproblem at each high-symmetry point to four independent subproblems on a quarter cell, and the resulting eigenfrequencies determine the stop-band boundaries, as shown below. Note that the Neumann/Dirichlet conditions replace the Bloch conditions only on the pair of faces whose phase factor varies along the considered segment; on the perpendicular pair of faces, the Bloch phase factor is fixed at its endpoint value $\lambda = +1$ (periodic) or $\lambda = -1$ (anti-periodic). For instance, along $\Gamma X$ the faces normal to y carry sound-hard or sound-soft conditions, while the faces normal to x remain Floquet-periodic with $\lambda_1$ fixed; along $XM$ the roles interchange, with $\lambda_1 = -1$ (anti-periodicity in x) held fixed. A sketch of the proposed approach is shown in Figure~\ref{fig:approach}.

The (anti-)periodic boundary conditions at high-symmetry points do not yet correspond to the standard Neumann (sound-hard) or Dirichlet (sound-soft) conditions of a closed cavity. To make this connection, an additional property is required, such that the material distribution must be invariant under the mirror reflections of the lattice point group. For the square lattice with the unit cell centered at the origin, the relevant mirrors are $\sigma_x\colon (x,y) \mapsto (-x,y)$ and $\sigma_y\colon (x,y) \mapsto (x,-y)$. It is assumed that $\sigma(\sigma_x\mathbf{r}) = \sigma(\mathbf{r})$ and $\sigma(\sigma_y\mathbf{r}) = \sigma(\mathbf{r})$, and identically for~$\tau$. Both a centered circular inclusion and a $2 \times 2$ checkerboard pattern satisfy this requirement, while an off-center inclusion does not.

When the mirror~$\sigma_x$ is a symmetry, it commutes with the operator in Eq.~\eqref{eq:wave} and every eigenstate can be decomposed into components of definite parity: $p = p_{\mathrm{e}} + p_{\mathrm{o}}$ with $p_{\mathrm{e}}(-x,y) = p_{\mathrm{e}}(x,y)$ and $p_{\mathrm{o}}(-x,y) = -p_{\mathrm{o}}(x,y)$. Both components satisfy Eq.~\eqref{eq:wave} individually, and the boundary conditions they inherit can be determined. Consider a point with $\lambda_1 = -1$ (anti-periodicity in~$x$). Assuming the evaluation at the right cell face $x = a/2$ for the even component, the condition $p_{\mathrm{e}}(-x,y) = p_{\mathrm{e}}(x,y)$ combined with anti-periodicity $p_{\mathrm{e}}(x+a,y) = -p_{\mathrm{e}}(x,y)$ gives,

\begin{equation}
  p_{\mathrm{e}}(a/2,\,y)
  = p_{\mathrm{e}}(-a/2,\,y)
  = -p_{\mathrm{e}}(a/2,\,y),
\label{eq:dirichlet_proof}
\end{equation}

where the first equality uses evenness and the second uses anti-periodicity. The latter forces $p_{\mathrm{e}}(a/2,y) = 0$, implying that the even mode satisfies the Dirichlet condition on the $x$-normal cell faces. For the odd component, the oddness implies $p_{\mathrm{o}}(0,y) = 0$ as the field vanishes at the cell center. Its normal derivative $\partial_x p_{\mathrm{o}}$ is even in~$x$, and an identical chain of reasoning yields

\begin{equation}
  \frac{\partial}{\partial x} p_{\mathrm{o}}(a/2,\,y) = 0\,,
\label{eq:neumann_proof}
\end{equation}

meaning that the odd mode satisfies the Neumann condition on the same faces. When $\lambda_1 = +1$ (periodicity in~$x$), the argument reverses: the even mode satisfies the Neumann condition, and the odd mode satisfies the Dirichlet condition. The same decomposition applies independently in~$y$ via the mirror~$\sigma_y$. Therefore, the complete spectrum at any high-symmetry point splits into four independent sub-problems on the quarter-cell $[0,\,a/2]^2$, labeled by the parity pair $(\pi_x,\pi_y) \in \{+,-\}^2$, each with Neumann or Dirichlet boundary conditions on each face as dictated by the combination of~$\lambda_j$ and~$\pi_j$ through the mechanism of Eqs.~\eqref{eq:dirichlet_proof} and~\eqref{eq:neumann_proof}. This decomposition is exact, following directly from the representation theory of the little group~$G_{\mathbf{k}_*}$ at the high-symmetry points. No perturbative or approximate step is involved.

We now turn to the spectral structure along a segment of the Brillouin zone boundary connecting two vertices $\mathbf{k}_A$ and~$\mathbf{k}_B$. Along the $\Gamma$X interval, the wavevector is $\mathbf{k}(t) = (t\pi/a,\,0)$ with $t \in [0,1]$. The mirror~$\sigma_y$ is preserved for all~$t$ (it maps $k_y \to -k_y$ and $k_y = 0$ throughout), so every band carries a conserved $\sigma_y$-parity label along this segment. The mirror~$\sigma_x$, however, is broken for $0 < t < 1$ because it maps $k_x \to -k_x$; the decomposition into even and odd sectors in~$x$ is therefore available only at the endpoints. Within each $\sigma_y$-parity sector, the eigenfrequencies $\omega_n(\mathbf{k}(t))$ are smooth functions of~$t$ and obey the non-crossing rule: bands of the same parity undergo avoided crossings. In contrast, bands of opposite parity may cross freely. As a consequence, the eigenvalue index~$n$ within each sector, and hence in the global ordering, is preserved from~$\mathbf{k}_A$ to~$\mathbf{k}_B$ continuously.

Let $H_1 \leq H_2 \leq H_3 \leq \cdots$ denote the eigenfrequencies at~$\mathbf{k}_A$ and $S_1 \leq S_2 \leq S_3 \leq \cdots$ those at~$\mathbf{k}_B$, both sorted in ascending order. The $n$-th band is a continuous curve connecting~$H_n$ to~$S_n$. If this curve is monotone in~$t$, meaning that the band attains its extrema only at the endpoints, then the band occupies exactly the frequency interval $[\min(H_n, S_n),\,\max(H_n, S_n)]$, and the stop-band between bands~$n$ and $n+1$ is the interval
\begin{equation}
  \Delta_n
  = \bigl[\,\max(H_n,\,S_n),\;\min(H_{n+1},\,S_{n+1})\,\bigr],
\label{eq:pairing}
\end{equation}
which is non-empty whenever 
\begin{equation}
    \max(H_n, S_n) < \min(H_{n+1}, S_{n+1}).
\end{equation}
A complete (omnidirectional) stop-band is obtained as the intersection of the partial gaps over all intervals of the irreducible Brillouin zone boundary.

Monotonicity of a band is not guaranteed in general; it fails when an avoided crossing in the interior of the segment causes the band to develop a local extreme that exceeds one of the endpoint values. The strength of such an avoided crossing is controlled by the coupling matrix element between the interacting modes, which in turn depends on how closely the modes are spaced in frequency and how strongly the perturbation $\delta\hat{H}(\mathbf{k})$ mixes them. A useful criterion is the ratio of the spectral isolation~$\delta_n$ to the bandwidth $g_n = |H_n - S_n|$, where $\delta_n$ is the minimum frequency distance from~$H_n$ or~$S_n$ to the nearest same-parity eigenfrequency at either endpoint. When $\delta_n \gg g_n$, the band is spectrally isolated, the avoided crossings are weak, and the band is monotone, which implies that Eq.~\eqref{eq:pairing} is exact. When $\delta_n \lesssim g_n$, strong mixing occurs, the band develops interior extrema, and the pairing rule becomes approximate. The latter explains the numerical observation that Eq.~\eqref{eq:pairing} reproduces the Floquet gaps to within numerical precision for the lowest bands and along the interval XM (where $k_x = \pi/a$ is fixed and the perturbation is quasi-one-dimensional). In contrast, higher bands along $\Gamma$X show progressively larger deviations.

The theoretical structure described above relies on two separate ingredients. The first is the existence of Brillouin zone points at which all Bloch phase factors satisfy $\lambda_j^2 = 1$, which is a property of the Bravais lattice alone, independent of the cell content. It ensures that the eigenstates at these points are standing waves with real boundary conditions. The second is the mirror symmetry of the unit cell, which splits the standing-wave eigenproblem into Neumann and Dirichlet sectors through the parity argument. Without mirror symmetry, the eigenstates at a high-symmetry point remain standing waves but do not decompose into pure even and odd components; the boundary conditions become mixed, and the clean pairing with cavity resonances is lost. The point group of the unit cell need not be the full~$C_{4v}$ of the square lattice; it suffices for it to contain the mirrors~$\sigma_x$ and~$\sigma_y$ that are normal to the cell faces. This condition is satisfied by any material distribution that is symmetric about both the $x$- and $y$-axes through the cell center.

The construction extends naturally to any Bravais lattice whose high-symmetry points are half-reciprocal-lattice vectors

\begin{equation}
    \mathbf{k}_* = \frac{1}{2}\sum_j n_j\mathbf{b}_j
\end{equation}

with $n_j \in \{0,1\}$, since these are exactly the points where $\lambda_j = e^{i\pi n_j} = \pm 1$. The square and rectangular lattices fall into this category. The hexagonal lattice, however, has a high-symmetry point~K at which $\lambda_j = e^{i2\pi/3}$, a cube root of unity. The Bloch boundary condition at~K does not square to the identity; it generates a cyclic group of order three rather than two, so the eigenspace decomposes into three sectors labeled by the three irreducible representations of~$\mathbb{Z}_3$, rather than two parity sectors. The Neumann/Dirichlet identification does not apply at such points, and the pairing rule in Eq.~\eqref{eq:pairing} cannot be used for segments terminating at~K without a generalization of the boundary-condition correspondence.

\subsection{Computational methods}
\label{sec:methods}

\paragraph{Plane-wave expansion.} Prior to the analysis of the arrays of Helmholtz resonators, the proposed method is validated on a canonical phononic crystal amenable to semi-analytical treatment, namely the square lattice of cylindrical inclusions in an elastic host, whose band structure was first computed within the plane-wave expansion in Ref.~\cite{kushwaha1994theory} and which has since served as a standard benchmark \cite{cao2004convergence}. Specifically, the band structure of a two-dimensional square-lattice phononic crystal composed of circular lead inclusions in an epoxy host is computed using the PWE. In this formulation, the spatially periodic inverse density $\sigma(\mathbf{r}) = 1/\rho(\mathbf{r})$ and the inverse bulk modulus $\tau(\mathbf{r}) = 1/\bigl(\rho(\mathbf{r})\, c(\mathbf{r})^2\bigr)$ are expanded in Fourier series over reciprocal lattice vectors~$\mathbf{G}$. For circular inclusions, the Fourier coefficients are expressed in closed form through the first-order Bessel function of the first kind. Truncating the expansion at~$N$ plane waves per spatial direction yields a generalized eigenvalue problem of dimension~$N^2$,

\begin{equation}
  \mathbf{A}(\mathbf{k})\,\mathbf{p} = \omega^2\,\mathbf{B}\,\mathbf{p},
  \label{eq:pwe}
\end{equation}

where $A_{ij}(\mathbf{k}) = \hat{\sigma}(\mathbf{G}_i - \mathbf{G}_j)\,(\mathbf{k}+\mathbf{G}_i)\cdot(\mathbf{k}+\mathbf{G}_j)$ and $B_{ij} = \hat{\tau}(\mathbf{G}_i - \mathbf{G}_j)$. The convergence of the eigenfrequencies with~$N$ is monitored, and $N = 17$ plane waves per direction ($289$~total) are found to provide sufficient accuracy for the first ten bands across the entire irreducible Brillouin zone.

The PWE framework has two roles here. First, it provides an independent, mesh-free reference solution against which the stop-band reconstruction procedure can be benchmarked, without ambiguities arising from finite-element discretization. Second, because the eigenvalue problem at any prescribed wavevector~$\mathbf{k}$ reduces to a standard matrix computation, the eigenfrequencies at the high-symmetry points $\Gamma$, X, and~M are obtained directly, which is the input required by the reconstruction algorithm described above. PWE thus allows the pairing rule in Eq.~\eqref{eq:pairing} to be assessed directly by comparison with the full Floquet dispersion diagram computed along the $\Gamma$--X--M--$\Gamma$ path.

\paragraph{Finite-element analysis.} The structures consisting of paired Helmholtz resonators are analyzed via finite-element calculations using  Comsol Multiphysics and the ``Pressure Acoustics, Frequency Domain'' physics in particular. In this case, the maximum element size is set to $1/10$ of the smallest wavelength in the considered spectral range $0 - 2500$~Hz.
To reconstruct the pass- and stop-bands of periodic structures, first, eigenmodes at high-symmetry points are calculated for unit cells with sound-hard and sound-soft boundaries. The first pass-band (i.e., at the lowest frequency) is located between the first pair of eigenmodes within the corresponding interval of the Brillouin zone. Then, knowing that each mode is the boundary between pass- and stop-bands, the band structure can be reconstructed. Since it is known that pass- and stop-bands are formed by modes of unit cells with different boundaries, the corresponding check is also provided. For that, the type of each mode is compared with the previous one. If they are the same, the ambiguity condition is assessed by averaging the field along the boundaries of the unit cell and along the mirror-symmetry axes. Then, if the condition is satisfied, the mode type can be changed. For example, if the acoustic field of a cell with sound hard boundaries is characterized by a node along the mirror symmetry line, it will be equivalent to the field of the unit cell with sound soft boundaries (see Figure~\ref{fig:ambiguity}).

\section{Results}
 \label{sec:results}
 
The decomposition developed in Section~\ref{sec:theory} imposes symmetry requirements on the \emph{unit cell} rather than on the scatterers it contains. What matters is that the material distribution is invariant under the mirrors~$\sigma_x$ and~$\sigma_y$ of the cell, a condition that constrains the placement and orientation of inclusions with respect to the cell center, but places no explicit restriction on their internal geometry. A complex meta-atom, such as a pair of coupled C-shaped Helmholtz resonators, satisfies the requirement in the same way as a simple circular cylinder, provided the unit cell as a whole possesses the necessary mirror symmetries. In other words, it is the symmetry of the boundary conditions on the cell faces that enables the standing-wave decomposition into Neumann and Dirichlet sectors, and the complexity of the interior wave field is immaterial.

To demonstrate this universality, the proposed method is applied to periodic systems of different physical nature with symmetric unit cells. First, in Section~\ref{sec:pwe_results}, a phononic crystal composed of lead cylinders in an epoxy matrix is analyzed using the PWE. The high symmetry and analytical tractability of this configuration allow the reconstruction procedure to be benchmarked against an independent, mesh-free reference solution, thereby isolating any discrepancies attributable to the pairing rule itself from those of numerical origin. Second, in Section~\ref{sec:HR_results}, the method is applied to a lattice of coupled Helmholtz resonators, a structure whose subwavelength resonances, narrow necks, and rectangular periodicity bear little resemblance to the Bragg-scattering physics of the first example.

\subsection{Phononic crystal: lead cylinders in epoxy}
\label{sec:pwe_results}

\begin{figure*}[htbp!]
    \centering
    \includegraphics[width=0.95\linewidth]{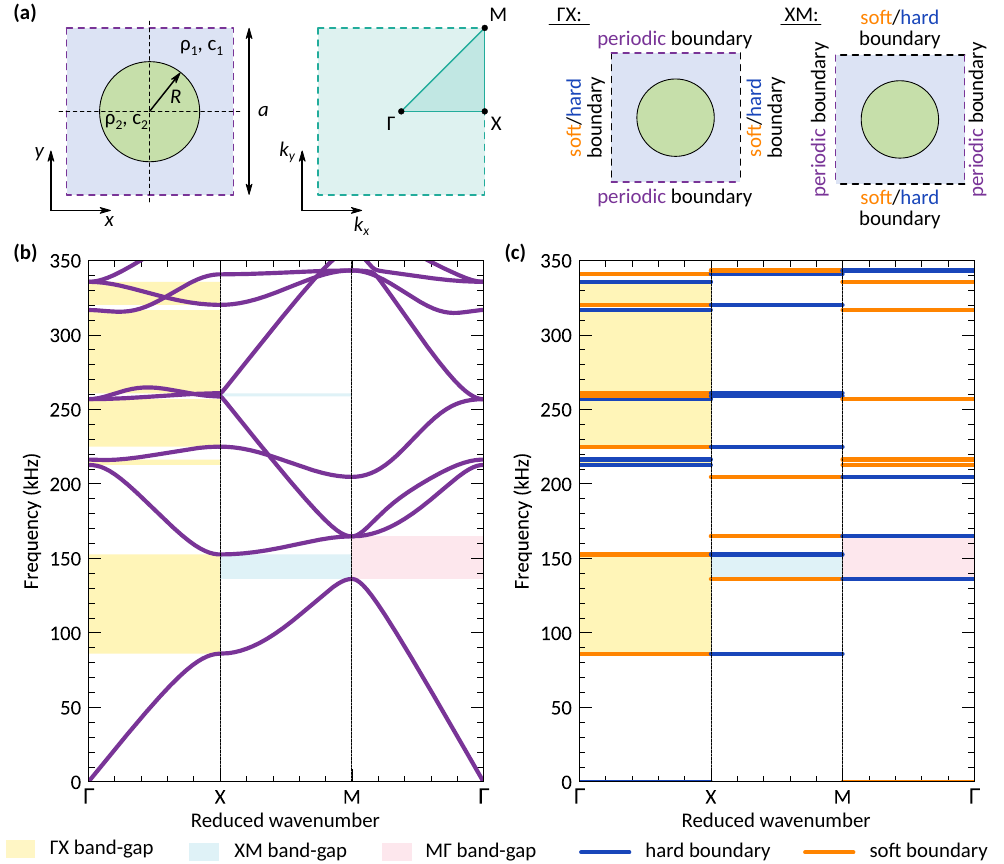}
    \caption{
    \textbf{Pass- and stop-bands of a phononic crystal of lead cylinders in an epoxy matrix.} 
    (a) Schematic of the square unit cell, the first Brillouin zone, and the boundary conditions used for the reconstruction. (b) Band structure with Floquet-periodic boundaries; shaded areas indicate band-gaps. (c) Mode structure for sound-hard (blue) and sound-soft (orange) boundaries.
    (d) Mode structure obtained for the unit cell with sound hard (blue lines) and sound soft boundaries (orange lines). Shaded areas on panels (b) and (c) indicate band-gaps, such that yellow color corresponds to $\Gamma$X band-gaps, light-blue to XM band-gaps, and pink to M$\Gamma$ band-gaps. By the $C_{4v}$ symmetry of the cell, the $\Gamma X$ results apply verbatim to the $\Gamma Y$ direction.
    }
    \label{fig:pwe}
\end{figure*}

As a first validation example, a two-dimensional phononic crystal consisting of lead cylinders periodically arranged in an epoxy matrix is considered [see Fig.~\ref{fig:pwe}(a)]. The unit cell is square with a side length $a = \SI{10}{\milli\meter}$, and each inclusion is a cylinder of radius $R = \SI{4}{\milli\meter}$ centered at the origin, corresponding to a filling fraction $f = \pi R^2 / a^2 \approx 0.503$. The epoxy (background) is characterized by the density $\rho_1 = \SI{1180}{\kilogram\per\cubic\meter}$ and the speed of sound $c_1 = \SI{2540}{\meter\per\second}$, while the density and speed of sound in lead (inclusion) are $\rho_2 = \SI{11340}{\kilogram\per\cubic\meter}$, $c_2 = \SI{2160}{\meter\per\second}$. The large contrast in both density and sound speed between the two constituents is known to favor the formation of wide Bragg-type stop-bands. This configuration is a classical test case for band-structure computations~\cite{kushwaha1994theory,cao2004convergence}, which makes it suitable for benchmarking the reconstruction procedure rather than for claiming new physics.

The first Brillouin zone of the square lattice is bounded by the high-symmetry points $\Gamma = (0,\,0)$, $\mathrm{X} = (\pi/a,\,0)$ and $\mathrm{M} = (\pi/a,\,\pi/a)$; the irreducible zone boundary is traced along the path $\Gamma$-X-M-$\Gamma$. At these points, the Bloch phase factors take values $\lambda_j \in \{+1,\,-1\}$, and the eigenstates reduce to standing waves whose symmetry properties are fully characterized by the mirrors~$\sigma_x$ and~$\sigma_y$ of the unit cell. Each segment of the irreducible Brillouin zone boundary probes a specific propagation direction: $\Gamma X$ corresponds to plane waves traveling along $x$, $XM$ to waves with fixed $k_X = \pi/a$ and varying $k_y$, and $M \Gamma$ to propagation along the diagonal. The shaded partial gaps, therefore, prohibit propagation only in the corresponding direction, while the omnidirectional stop-band is their intersection. For the square unit cell, whose point group $C_{4v}$ contains the diagonal mirror mapping $\Gamma X$ onto $\Gamma Y$, the partial stop-bands for x- and y-propagating waves coincide identically, so the $\Gamma Y$ segment need not be computed; this equivalence does not hold for the rectangular lattice of Section~\ref{sec:HR_results}, where the $x-$ and $y-$direction gaps differ and must be evaluated separately.

As established in Section~\ref{sec:theory}, each eigenstate can then be classified according to its behavior on the cell boundary. Modes that are even under the relevant mirror satisfy the sound-hard (Neumann) condition, whereas modes that are odd under the same mirror satisfy the sound-soft (Dirichlet) condition. The boundary conditions applicable to each segment of the Brillouin zone boundary are illustrated schematically in Fig.~\ref{fig:pwe}(b). For the $\Gamma$X interval, $\sigma_y$~is preserved throughout; accordingly, the boundaries normal to the $y$-axis carry sound-hard or sound-soft conditions, while those normal to~$x$ remain Floquet-periodic. For the XM interval, the roles are interchanged: Floquet periodicity is imposed along~$y$, and sound-hard or sound-soft conditions along~$x$.

Figure~\ref{fig:pwe}(c) shows the band structure obtained by solving the full Floquet eigenvalue problem along $\Gamma$--X--M--$\Gamma$. The dispersion curves (solid lines) reveal a complex modal landscape with multiple stop-bands, shaded according to the Brillouin zone segment in which they occur: yellow for the $\Gamma$X, light blue for XM, and pink for M$\Gamma$ intervals. A complete (omnidirectional) stop-band is identified in the frequency range where the partial gaps overlap across all three segments. Figure~\ref{fig:pwe}(d) presents the result of the corresponding reconstruction procedure. Rather than computing the full dispersion diagram along several hundred wavevectors, only the eigenfrequencies at the three high-symmetry points are required. These are displayed as horizontal lines spanning the corresponding Brillouin zone segment: blue lines denote eigenfrequencies obtained with sound-hard boundary conditions, and orange lines indicate those obtained with sound-soft conditions. For each segment, the eigenfrequencies at its two endpoints are sorted in ascending order and paired by index: the $n$-th band is assigned following Eq.~\eqref{eq:pairing}. A partial stop-band between the $n$-th and $(n+1)$-th bands exists whenever $\max(H_n,\, S_n) < \min(H_{n+1},\, S_{n+1})$. The resulting stop-band intervals, shaded in yellow ($\Gamma$X), light blue (XM), and pink (M$\Gamma$), are seen to coincide with those obtained from the full Floquet calculation in Fig.~\ref{fig:pwe}(c).

Agreement is closest for the lowest bands and for the XM interval, where the perturbation along the path is quasi-one-dimensional, and the monotonicity assumption underlying Eq.~\eqref{eq:pairing} is well satisfied. For this configuration, all reconstructed gap boundaries agree with the Floquet reference to within 1\%, including the higher-order bands along $\Gamma X$ where avoided crossings slightly perturb band monotonicity.

\subsection{Complex geometry: Helmholtz resonators}
\label{sec:HR_results}

\begin{figure*}[ht!]
    \centering
    \includegraphics[width=0.9\linewidth]{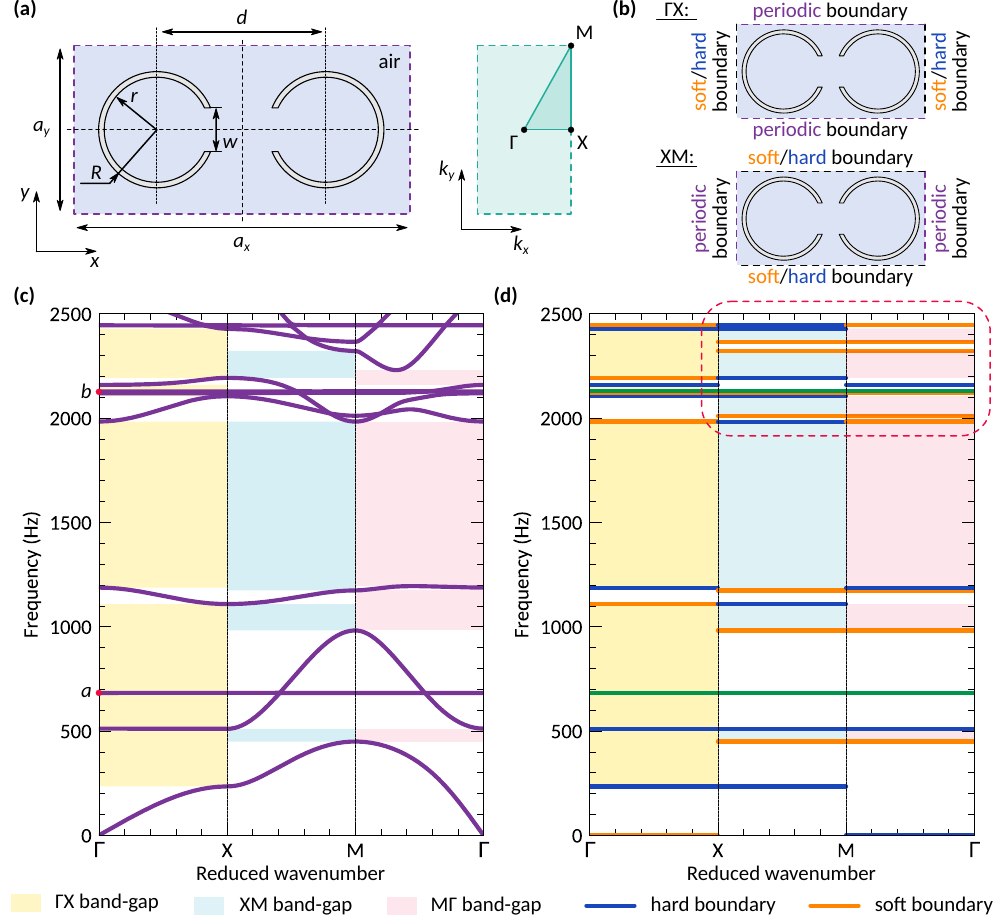}
    \caption{
    \textbf{Pass- and stop-bands of a periodic structure consisting of coupled Helmholtz resonators.} 
    (a) Schematic illustration of the considered rectangular unit cell and the corresponding first Brillouin zone. (b) Boundary conditions utilized for the reconstruction of pass- and stop-bands.
    (c) Band structure obtained for the case when all boundaries of the unit cell are Floquet-periodic. Shaded areas indicate band-gaps. 
    (d) Mode structure obtained for the unit cell with sound hard (blue lines) and sound soft boundaries (orange lines). Green lines correspond to bound states in the continuum. Shaded areas indicate band-gaps. The red dashed rectangle highlights the area where the estimation of band-gap location and width becomes inaccurate.
    }
    \label{fig:coupled_HR_modes}
\end{figure*}

As an example of an acoustic periodic system, a structure consisting of coupled C-shaped Helmholtz resonators is considered [see Fig.~\ref{fig:coupled_HR_modes}(a)]. Specifically, the pairs of resonators are arranged in a rectangular lattice, meaning that the first irreducible Brillouin zone is defined by the $\Gamma$, X, and M points of high symmetry. 

The resonators are characterized by the outer radius $R = 53$~mm, the inner radius $r = 48$~mm, the slit width $w = 40$~mm, and the distance between the resonators $d = 120$~mm. The width of the unit cell is $a_x = 240$~mm, and the height is $a_y = 120$~mm. It is assumed that all resonator boundaries are sound-hard, and the wall material is therefore irrelevant. This system was chosen because periodic structures based on this type of meta-atoms are characterized by broad stop-bands~\cite{krasikova2023metahouse,krasikova2024broadband}; the band structure also exhibits flat bands corresponding to bound states in the continuum (BIC) (see Fig.~\ref{fig:coupled_HR_modes}(c) and Section~\ref{sec:conclusion}).

In the considered acoustic system, where the wave propagates in air, the class-consistent conditions~\cite{hvatov2015free, sorokin2022finite} reduce to sound-soft and sound-hard boundaries (see Fig.~\ref{fig:coupled_HR_modes}(b)), for which the pressure or the normal velocity component vanishes, respectively.

Figure~\ref{fig:coupled_HR_modes}(c) demonstrates the spectral position of the eigenmodes for the case of unit cells with different boundary conditions. Specifically, the modes in the $\Gamma$X interval are obtained for the case when boundaries along the $y$-axis are sound soft/hard, and the boundaries along the $x$-axis are Floquet-periodic. For the XM interval, these conditions are interchanged, such that Floquet periodic boundaries are along the $y$-axis and sound soft/hard ones are along the $x$-axis. Then, the modes in the M$\Gamma$ interval are reconstructed using those from $\Gamma$X and XM intervals. 
The locations of the pass and stop-bands, defined by the algorithm described in Methods, coincide with the corresponding ranges obtained when all boundaries are Floquet-periodic.

\begin{figure*}[htbp!]
    \centering
    \includegraphics[width=0.9\linewidth]{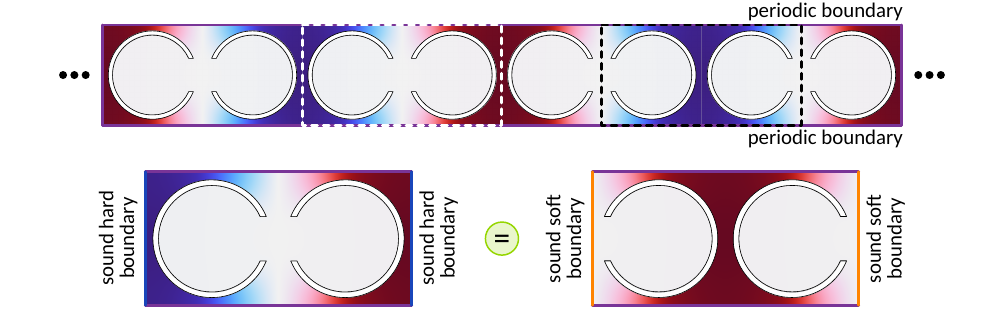}
    \caption{Illustration of the ambiguity in the unit cell choice. Periodicity of the structure implies that a unit cell can be chosen in different ways. Depending on this selection, the field distributions can be characterized by nodes or antinodes at the unit cell boundaries. Hence, the same field distribution can be obtained using both sound-hard and sound-soft boundaries, provided the unit cell origin is displaced accordingly.  }
    \label{fig:ambiguity}
\end{figure*}

The periodicity of the structure implies ambiguity in the choice of the unit cell. As established for elastic periodic waveguides in \cite{sorokin2022finite}, a shift of the unit cell origin by half a period leaves the eigenfrequency spectrum unchanged while interchanging the boundary-condition class of every eigenfrequency. In the present acoustic setting, this is illustrated in Fig.~\ref{fig:ambiguity}: the field distribution in the cell with sound-hard boundaries coincides with that of the half-period-shifted cell with sound-soft boundaries. A consequence specific to the structures considered here is that BIC modes appear at identical frequencies under both boundary types (see Section~\ref{sec:conclusion} for further discussion).

The reconstruction becomes inaccurate when modes intersect within a Brillouin zone interval. In the considered case, such a situation can be observed in the X--M--$\Gamma$ region at frequencies above $2000$~Hz, as indicated in Fig.~\ref{fig:coupled_HR_modes}(d).

\subsection{Photonic crystal}

\begin{figure*}[htbp!]
    \centering
    \includegraphics[width=0.9\linewidth]{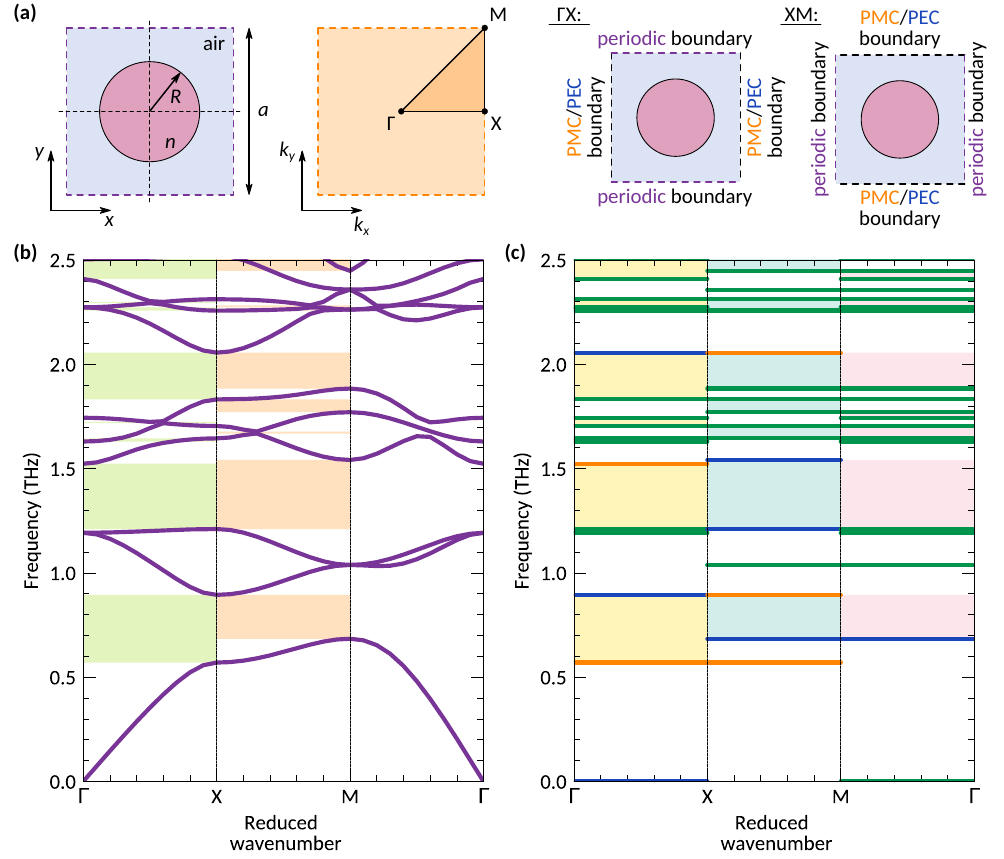}
    \caption{\textbf{Pass- and stop-bands of a photonic crystal.} (a) Schematics of the considered square unit cell consisting of a cylinder made of an isotropic uniform material. The corresponding reciprocal unit cell is then also a square. In order to reconstruct the positions of the pass- and stop-bands, the perfect electric conductor (PEC) and perfect magnetic conductor (PMC) boundary conditions are applied. (b) Band structure of the photonic crystal, such that the unit cell size is $a = \SI{100}{\micro\meter}$, the radius of the cylinder is $R = \SI{30}{\micro\meter}$ and its refractive index is $n=3.5$. Shaded areas indicate the stop-bands. (c) Mode structure obtained for the unit cell with the PEC (blue lines) and PMC (orange lines) boundary conditions. Shaded areas indicate band-gaps.}
    \label{fig:phc}
\end{figure*}

As a further demonstration of generality, the method is applied to a photonic crystal of dielectric cylinders in a square lattice. The results are shown in Fig.~\ref{fig:phc}. Specifically, the considered square unit cell consists of uniform cylinders made of a material with the refractive index $n = 3.5$ [see Fig.~\ref{fig:phc}(a)]. The side length of the cell is $a = \SI{100}{\micro\meter}$ while the radius of the cylinder is $R = \SI{30}{\micro\meter}$. Such periodic structures exhibit band-gaps, which is also shown in Fig.~\ref{fig:phc}(b). As in the acoustic case, pass- and stop-bands can be reconstructed using perfect electric (PEC) and perfect magnetic (PMC) boundary conditions. As shown in Fig.~\ref{fig:phc}(c), the positions of band-gaps correspond well to the gaps in the Floquet band diagram.

\section{Discussion and conclusions}
\label{sec:conclusion}

The mirror symmetries of a periodic unit cell have been shown to enable direct reconstruction of stop-band boundaries from a handful of unit-cell eigenvalue problems, without computing the full Floquet dispersion diagram. The key theoretical result is that, for unit cells possessing mirror symmetries $\sigma_x$ and $\sigma_y$, the eigenstates at the high-symmetry vertices of the Brillouin zone ($\Gamma$, X, and M for square and rectangular lattices) decompose into four independent parity sectors. Each sector satisfies Neumann or Dirichlet boundary conditions on the cell faces, as dictated by the combination of the Bloch phase factor $\lambda_j \in \{+1, -1\}$ and the eigenstate parity $\pi_j \in \{+, -\}$. This decomposition is exact and follows from the representation theory of the little group $G_{\mathbf{k}^*}$ at each high-symmetry point.

By sorting the eigenfrequencies at the endpoints of each Brillouin zone interval and pairing them by index, the pass-bands and stop-bands are reconstructed through the rule Eq.~\eqref{eq:pairing}. The validity of this reconstruction has been confirmed on two physically distinct configurations. For the phononic crystal consisting of lead cylinders in an epoxy matrix (the phononic crystal of Section~\ref{sec:pwe_results}), the plane-wave expansion provides a mesh-free reference, and the reconstructed gap boundaries deviate from the full Floquet calculation by less than $1\%$ across all identified stop-bands. For the lattice of coupled C-shaped Helmholtz resonators, finite-element calculations in COMSOL Multiphysics demonstrate a faithful reproduction of the stop-band map, despite the subwavelength resonant nature of the gaps and the geometric complexity of the unit cell interior.

Three aspects of the method warrant comment. First, the computational savings are substantial, especially for finite-element models with fine meshes: the reconstruction requires eigenvalue solutions at only three discrete wavevectors (or, equivalently, six quarter-cell problems per vertex when the symmetry decomposition is exploited), whereas the full Floquet calculation demands sweeping the entire Brillouin zone boundary, which is here discretized into $601$ wavevector samples. Second, the method depends solely on the lattice symmetry and is independent of the scattering mechanism (Bragg or local resonance), the frequency regime, and the geometric complexity of the meta-atom. The only requirement is that the material distribution within the unit cell be invariant under the mirrors normal to the cell faces, a condition satisfied by many practical designs.

The pairing rule is exact as long as each band is monotone along the segment, which is guaranteed when the band is spectrally isolated ($\delta_n \gg g_n$, Section~\ref{sec:theory}). When this condition is violated, attenuation intervals arise whose boundaries do not coincide with unit-cell eigenfrequencies. Such intervals are the two-dimensional counterpart of the 'flutter-type' stop-bands recently identified in multi-coupled periodic waveguides \cite{firouzi2026classification}, which, in contrast to the conventional 'divergence-type' gaps bounded by standing-wave band edges, originate from the interaction of wave pairs in the interior of the segment and are not detectable from the unit-cell spectrum alone. In the present examples, this situation is confined to the $X-M-\Gamma$ region above 2000 Hz for the resonator lattice (Fig.~\ref{fig:coupled_HR_modes}) and to higher-order bands along $\Gamma X$ for the phononic crystal.

A by-product of the analysis concerns bound states in the continuum (BIC). The flat bands observed in the Helmholtz resonator band structure correspond to modes whose fields vanish along both the unit-cell boundaries and the mirror-symmetry axes, implying $p = 0$ at these locations. Such modes satisfy both sound-hard and sound-soft boundary conditions simultaneously, a consequence of the class-interchange property established in \cite{sorokin2022finite}, which explains their appearance at identical frequencies for both boundary types. The proposed method provides a basis for directly identifying BIC spectral positions from unit-cell eigenfrequency analysis, without a full band-structure calculation.

The present formulation applies to Bravais lattices whose zone vertices are half-reciprocal-lattice vectors, i.e., points where every Bloch phase factor equals $\pm 1$. Square and rectangular lattices fall into this category. The limitation of the method to Bravais lattices with $\lambda_j \in \{\pm 1\}$
at all zone vertices (Section~\ref{sec:theory}) excludes the hexagonal lattice, where extension to $\mathbb{Z}_3$-valued phases remains an open problem.

\section*{Funding}
M.Kr. and A.P. acknowledge the financial support from the Russian Science Foundation (\href{https://rscf.ru/en/project/24-21-00275/}{project \textnumero 24-21-00275}).

\section*{Author contributions}
A.H. developed the theoretical description, M.K. and A.P. performed numerical calculations, and S.M. supervised the work. All authors contributed to the analysis of the results and preparation of the manuscript.

\section*{Declaration of competing interest}
The authors declare no conflicts of interest.

\section*{Acknowledgments}
The authors thank Sergey Sorokin for fruitful discussions and useful comments. M.K. and A.P. thank Sergey Krasikov for the discussions and suggestions.

\section*{Data availability}
No experimental data were generated. All material data and required methods to reproduce the results are available in the paper.

\bibliographystyle{unsrt}
\bibliography{references.bib}

\end{document}